# Meta-Learning Initializations for Interactive Medical Image Registration

Zachary M. C. Baum, Yipeng Hu, and Dean C. Barratt

***Abstract*—We present a meta-learning framework for interactive medical image registration. Our proposed framework comprises three components: a learning-based medical image registration algorithm, a form of user interaction that refines registration at inference, and a meta-learning protocol that learns a rapidly adaptable network initialization. This paper describes a specific algorithm that implements the registration, interaction and meta-learning protocol for our exemplar clinical application: registration of magnetic resonance (MR) imaging to interactively acquired, sparsely-sampled transrectal ultrasound (TRUS) images. Our approach obtains comparable registration error (4.26 mm) to the best-performing non-interactive learning-based 3D-to-3D method (3.97 mm) while requiring only a fraction of the data, and occurring in real-time during acquisition. Applying sparsely sampled data to non-interactive methods yields higher registration errors (6.26 mm), demonstrating the effectiveness of interactive MR-TRUS registration, which may be applied intraoperatively given the real-time nature of the adaptation process.***

## I. INTRODUCTION

### A. Medical Image Registration

IMAGE registration is a fundamental task in medical imaging research whereby correspondence is established between anatomical structures in paired images. Using methodologies from "classical" iterative registration algorithms, learning-based methods have been proposed. Learning-based methods have used different architectures, such as convolutional neural networks [1-2] and vision transformers [3], different training strategies, such as generative adversarial networks [4, 5], supervised [1, 4], unsupervised [2, 6-8] or reinforcement learning [9-11], or different transformation constraints, based on parametric splines [6], diffeomorphism [12] and biomechanics [13]. Semi-supervised learning [14], few-shot- and meta-learning [15-16], unsupervised contrastive learning [17], inference-time augmentation [16, 18], and amortized hyperparameter learning [19] methodologies have also been used to improve data efficiency and generalizability. For further discussion on these learning-based registration methods, readers are referred to recent systematic surveys [20-22].

### B. Interactive Machine Learning

For many clinical applications, registration errors may be identified or corrected by users, and integration of interactions into machine learning frameworks may assist in predicting more accurate solutions [23-26]. This integration is referred to as interactive machine learning (IML) [27]. Recently, IML-based methods for medical image analysis have focused on error correction in image segmentation. Most existing methods use simple interactions, such as user-defined bounding boxes as a guide for initial predictions [24] or 'scribbles' to indicate areas that should or should not be considered during refinement [23-24, 26]. Other methods may train multiple networks in tandem; one for segmentation and another for refinement [25].

Though the authors were unable to find existing IML-based methods for medical image registration in the literature, interaction has been utilized in classical methods. To improve the alignment of patient images, anatomical landmarks may be interactively selected [28] or acquired with spatially tracked intra-operative surgical instruments [29]. These methods are widely used and considered among the gold standard for medical image analysis methods.

### C. Gradient-Based Meta-Learning

Meta-learning [30-31] formalizes the commonly-applied fine-tuning intuition by iteratively learning to improve future performance on related tasks over multiple learning episodes. In particular, 'gradient-based' meta-learning approaches, such as model agnostic meta-learning (MAML) [32] and Reptile [33], learn adaptable initializations from gradients observed during learning episodes. Such algorithms are simple, learn quickly, and generalize well at test time with limited examples, as evidenced by their application in the medical imaging domain [15-16, 34-37]. Gradient-based methods have been used for domain-agnostic generalization and subsequent rapid-adaptation in image registration [15-16] and segmentation [34-37] on datasets of limited size from new domains.

Z.M.C. Baum is supported by the Natural Sciences and Engineering Research Council of Canada Postgraduate Scholarships-Doctoral Program, and the University College London Overseas and Graduate Research Scholarships. This research was funded in whole, or in part, by the Wellcome Trust [203145Z/16/Z]. This work is also supported by the International Alliance for Cancer Early Detection, an alliance between Cancer Research UK [C28070/A30912; C73666/A31378], Canary Center at Stanford University, the University of Cambridge, OHSU Knight Cancer Institute, University College London and the University of Manchester. For the purpose of Open Access, the author has applied a CC BY public copyright licence to any Author Accepted Manuscript version arising from this submission.

Zachary M. C. Baum, Yipeng Hu, and Dean C. Barratt are with the Wellcome/EPSRC Centre for Interventional and Surgical Sciences, University College London, London W1W 7TS, U.K., and the UCL Centre for Medical Image Computing, University College London, London W1W 7TS, U.K. (e-mail: zachary.baum.19@ucl.ac.uk; yipeng.hu@ucl.ac.uk; d.barratt@ucl.ac.uk).

Unlike most aforementioned examples, this work focuses on improving performance for individual tasks, formed by data that are varied by interactions. The simplicity in incorporating new data, efficiency in adaptation, and effectiveness in various computer vision and medical imaging applications are particularly desirable and motivate the use of meta-learning in formulating interactive registration. Other meta-learning methodologies [30] should be tested in future development.

### D. Contributions

We define a framework to meta-learn network initializations for interactive image registration. This framework consists of three components: a learning-based medical image registration algorithm, a form of user interaction, which is easily simulated in training, to refine predictions at inference, and a gradient-based meta-learning protocol that learns a rapidly adaptable network initialization, by considering data variability due to interaction in individual patients as separate tasks.

To investigate the application of such a framework to clinical data, we register 3D magnetic resonance (MR) imaging volumes to a series of interactively-acquired sparse 2D transrectal ultrasound (TRUS) images for use in targeted prostate biopsy guidance. This exemplar application illustrates a clinical scenario in which real-time, interventional imaging, such as TRUS, is acquired interactively to iteratively refine the registration throughout a single acquisition of interventional imaging modality as it traverses the target anatomy. This work compares the accuracy of our proposed interactive registration method with alternative learning-based methods. We outline the key contributions in this work as follows:

- We provide a detailed description of our interactive meta-learning framework for medical image registration and describe how it may enable a range of useful applications.
- We introduce and describe the registration, interaction, and meta-learning strategy for our exemplar clinical application; volume-to-sparse registration of prostate MR to TRUS.
- We present rigorous validation experiments, comparing our method to various learning-based methods for prostate MR-TRUS registration, including variations to meta-learning parameters to assess their effects on the registration process.

## II. LEARNING-BASED INTERACTIVE IMAGE REGISTRATION

### A. Learning-based Image Registration

Learning-based registration may be categorized from an application perspective; network inputs may be unimodal, multimodal, inter-patient, or intra-patient – with each image bearing its own dimensionality [20], requiring different loss functions based on image similarity [6], label similarity [1], or some combination of the two [2]. Each image pair may encompass any number of anatomical sites of clinical interest, requiring a registration method to utilize different deformation models, commonly, rigid, affine, or deformable [20].

Given $N$ pairs of training source and target images, $\{x_n^{source}\}$ and $\{x_n^{target}\}$, and accompanying source and target labels, $\{l_n^{source}\}$ and $\{l_n^{target}\}$, respectively, where $n = 1, \ldots, N$, existing approaches predict the voxel correspondence or transformation $u_n^\phi = f^\phi(x_n^{source}, x_n^{target})$ using a registration network $f^\phi$ with network parameters or weights $\phi$. The training goal thus is minimizing an image and/or label loss function $\mathcal{L}_{sim}$ over $N$ training pairs, to obtain the optimal $\hat{\phi}$:

$$\hat{\phi} = \arg\min_\phi \sum_{n=1}^N [\mathcal{L}_{sim}(\phi) + \alpha^{def} \mathcal{L}_{def}(\phi)], \quad (1)$$

where $\mathcal{L}_{def}(\phi | x_n^{source}, x_n^{target}) = \mathcal{L}_{def}(f^\phi(x_n^{source}, x_n^{target}))$ provides regularization on the deformation smoothness $u_n^\phi$, weighted by $\alpha^{def}$. In general, the similarity-based loss can further combine a negative unsupervised image similarity function $\mathcal{L}_{sim}^{image}(x_n^{source}(u_n^\phi), x_n^{target})$, between the transformation-warped images $x_n^{source}(u_n^\phi)$ and the target images $x_n^{target}$, and a negative weak-supervision loss based on label similarity $\mathcal{L}_{sim}^{label}(l_n^{source}(u_n^\phi), l_n^{target})$, between the warped source labels $l_n^{source}(u_n^\phi)$ and the target labels $l_n^{target}$:

$$\mathcal{L}_{sim}(\phi | x_n^{source}, x_n^{target}, l_n^{source}, l_n^{target}) = $$
$$\alpha^{image} \mathcal{L}_{sim}^{image}(x_n^{source}(u_n^\phi), x_n^{target}) + $$
$$\alpha^{label} \mathcal{L}_{sim}^{label}(l_n^{source}(u_n^\phi), l_n^{target}), \quad (2)$$

where the general form contains hyperparameters $\alpha^{image}$ and $\alpha^{label}$ which may be set to zero to represent weakly supervised and unsupervised algorithms, respectively.

### B. Interaction for Image Registration

In general, the performance improvement seen in other interactive applications, such as the above-discussed interactive segmentation [34-37], may be expected from interactive registration. Other benefits, such as those related to expandability, and owing to the human-in-the-loop of machine learning models for registration applications are also important, but are considered out of the scope for this work.

To adapt existing learning-based registration methods to accept interactions, we must first define interactions that may be learned in training, and are feasible at test-time. We consider interaction to be any action taken by the user. Depending on application-specific needs, a combination of sequential interactions may best improve the registration.

This user-to-computer interaction may entail image re-acquisition or annotation of poorly aligned areas. Image re-acquisition may be local (i.e. one, or a few images) or global (i.e. entire image volume) when, for example, image quality is poor, or there has been patient motion. Local re-acquisition is pertinent when using real-time imaging modalities, such as ultrasound, that can be rapidly acquired.

We propose formulating image reacquisition and annotation additional labelled data, where the quantity and availability of labels or images may vary per application. In practice, interactions may be application-specific. The determination of use-cases for each interaction is considered out of scope for this work, though we provide a description and evaluation of the additional data interaction for ultrasound-guided prostate biopsy to illustrate a possible use of the proposed framework.

### C. Meta-Learning Interactive Initializations

In this work, we train registration networks in the inner loop of a meta-learning optimization to accept newly labelled data provided by interaction, while the network adaptability across subjects is optimized in an outer loop. For a given test subject, this enables the interactively-acquired data to adapt the trained registration network efficiently, before being used in inference.

## III. METHODS

### A. Images and Annotations as Interaction

We denote possible pairs of interactions sampled from the source and target images as $\{i_{mn}^{source}\}$ and $\{i_{mn}^{target}\}$, from $n$ training data. Each $n^{th}$ pair is also associated with $M_n$ interactions that are possible on image pair $n$, $m = 1, \ldots, M_n$. These time-agnostic interactions are represented as sets of interactively obtained images ($\{x_{mn}^{source}\}$ and $\{x_{mn}^{target}\}$) and annotations, in the form of segmentation labels, ($\{\ell_{mn}^{source}\}$ and $\{\ell_{mn}^{target}\}$), i.e. $i_{mn}^{source} = [(x_{mn}^{source})^T, (\ell_{mn}^{source})^T]^T$ and $i_{mn}^{target} = [(x_{mn}^{target})^T, (\ell_{mn}^{target})^T]^T$. For notational brevity, both images and annotations can include the previously available annotated data, for individual subject, therefore the interactions $\{i_{mn}^{source}\}$ and $\{i_{mn}^{target}\}$ are interchangeably used with interaction-updated source and target, respectively. A sequence of interactions may benefit from explicit sequential modelling; however this is considered out of scope of this work, where only a few steps of interaction is considered feasible in the application of interest.

This formulation does not distinguish between registrations which may have different initial image and annotation data from one without such initial registration, as they can be consistently represented by both the non-interactive registration formulation, described in Section II.A, and the interactive adaptation, described in Section III.B.

We note that not all the interactive image or annotation data need to be available or varying for a given interaction. We describe a sample of scenarios which demonstrate the versatility of interactive registration. Additionally, active learning methodologies [38] may appear similar in nature, and may be able to utilize similar scenarios for interactive learning in practice. Our application is developed and validated with respect to Scenario 4, a special case of Scenario 3. Though not tested, other scenarios are included for discussion purposes.

1. Successive user-defined image annotations improve the registration over multiple interactions, i.e. variable labels $\ell_{m=a,n}^{source} \neq \ell_{m=b,n}^{source}$ and $\ell_{m=a,n}^{target} \neq \ell_{m=b,n}^{target}$ but fixed images $x_{m=a,n}^{source} = x_{m=b,n}^{source}$ and $x_{m=a,n}^{target} = x_{m=b,n}^{target}$, when $a \neq b$.
2. An unsupervised learning algorithm, without initial labels $\ell_{m=0,n}^{source}$ and $\ell_{m=0,n}^{target}$, receives successive user-defined annotations to improve alignment, which requires the simulation of $\ell_{m>0,n}^{source}$ and $\ell_{m>0,n}^{target}$ during training.
3. An image-guidance application may have a fixed pre-operative image $x_{mn}^{source}$, but adds new intra-operative images, i.e. $x_{m=a,n}^{source} \neq x_{m=b,n}^{source}$ and $x_{m=a,n}^{target} = x_{m=b,n}^{target}$, when $a \neq b$. This application may use user-defined annotations on the pre- and intra-operative images, as in Scenario 1.
4. An ultrasound-guided prostate cancer application, such as that used in this work; similar to Scenario 3, but does not require new annotations on the source images, however, additional annotations on the target images may be acquired automatically using a segmentation network, i.e. using the generation of labelled intra-operative ultrasound images as the interaction, $x_{m=a,n}^{source} = x_{m=b,n}^{source}$, $\ell_{m=a,n}^{source} = \ell_{m=b,n}^{source}$, $x_{m=a,n}^{target} \neq x_{m=b,n}^{target}$ and $\ell_{m=a,n}^{target} \neq \ell_{m=b,n}^{target}$, when $a \neq b$.

### B. Meta-Learning for Interactive Registration

As the interaction data $\{i_{mn}^{source}\}$ and $\{i_{mn}^{target}\}$ are defined as images and annotations – in Section III.A – $\{x_{mn}^{source}\}$, $\{x_{mn}^{target}\}$, $\{\ell_{mn}^{source}\}$ and $\{\ell_{mn}^{target}\}$, they are consistent with the data used in the non-interactive registration formulation – in Section II.A – $\{x_n^{source}\}$, $\{x_n^{target}\}$, $\{l_n^{source}\}$ and $\{l_n^{target}\}$. We propose to formulate the training of an interactive registration network $f^{\tilde{\phi}}$ by adapting the optimization in Eq. (1) to a bi-level optimization [30, 38], therefore learning the interactive image registration becomes a meta-learning problem:

$$\tilde{\phi} = \arg\min_{\phi} \sum_{n=1}^{N} \sum_{m=1}^{M_n} [\mathcal{L}_{sim}^*(\phi^*(\phi)) + \alpha^{def} \mathcal{L}_{def}^*(\phi^*(\phi))], \quad (3)$$

$$s.t. \, \phi^* = \arg\min_{\phi} \sum_{n=1}^{N} \sum_{m=1}^{M_n} [\mathcal{L}_{sim}^*(\phi) + \alpha^{def} \mathcal{L}_{def}^*(\phi)], \quad (4)$$

where $\mathcal{L}_{sim}^*$ is obtained by substituting interaction in Eq. (2):

$$\mathcal{L}_{sim}^*(\phi) = \mathcal{L}_{sim}(\phi | x_{mn}^{source}, x_{mn}^{target}, \ell_{mn}^{source}, \ell_{mn}^{target}), \quad (5)$$

similarly, $\mathcal{L}_{def}^*(\phi) = \mathcal{L}_{def}(\phi | x_{mn}^{source}, x_{mn}^{target})$. $\mathcal{L}_{sim}^*(\phi^*(\phi))$ and $\mathcal{L}_{def}^*(\phi^*(\phi))$ denote the optimized functions of $\phi$, by optimized $\phi^*$ at the inner-level. $\phi^*$ hereinafter used for brevity.

It is noteworthy that, unlike the training defined in Eq. (1) which minimizes the expected loss over the $N$ pairs of training images, the task-specific inner-level Eq. (4) aims to minimize the expected loss over the $M_n$ samples of interactions. At the outer-level, Eq. (3), different $N$ pairs of images and annotation are usually sampled to learn the optimal network parameters, such that, at inference, the network $f^{\tilde{\phi}}$ can be adapted to new pairs of interactions $\{i_{m,test}^{source}\}$ and $\{i_{m,test}^{target}\}$, where $m = 1, \ldots, M_{test}$ and be generalized to this new *test* task, i.e. we define the training meta-tasks be $N$ different cases that need registration, rather than $M_n$ steps of interactions.

Such a meta-learning framework learns an initialization of network parameters $\tilde{\phi}$ which enables data-efficient adaptation to a new task at inference. The efficient adaptation means that registering a new pair of images $x_{test}^{source}$ and $x_{test}^{target}$ may only require a few $M_{test}$ steps of interaction, often constrained by human effort and time-critical applications.

### C. Gradient-Based Meta-Learning Algorithms for Network Initialization

Gradient-based meta-learning algorithms are applicable for training the proposed interactive registration and are comprised of the meta-learning and the meta-test phases. To start meta-training, the registration model is initialized with random weights. During each iteration of the outer-level loop, one task $(i_{mn}^{source}, i_{mn}^{target})_n$ is randomly sampled from the task set $\{(i_{mn}^{source}, i_{mn}^{target})_{n=1,\ldots,N}\}$ containing all possible tasks, with a set of $k$ interactions $\{(i_{mn}^{source}, i_{mn}^{target})_{n,m=1,\ldots,k}\}$ randomly sampled from this given task, to form an episode (Fig. 1). Each sampled task corresponds to a task-specific loss in Eq. (4). We define our meta-learning task as a pair of source and target images with their associated source and target annotations, from each subject. During each episode we undergo 'task-level learning' using stochastic gradient descent (SGD) or its variants, for $k$ SGD steps, the task-specific gradient $g_n^m(\phi)$ can be computed to update the network weights $\phi$:

$$\phi_m^* \leftarrow \phi - \beta^{task} \cdot g_n^m(\phi), \quad (6)$$

$$\text{where } g_n^m(\phi) = \frac{\partial}{\partial \phi}\big[\mathcal{L}_{sim}^*(\phi) + \alpha^{def}\mathcal{L}_{def}^*(\phi)\big], \quad (7)$$

and $\beta^{task}$ is the learning rate. After an episode of $k$ steps, a cross-task gradient $g_n(\phi^*)$ is used to update the network weights at the outer-level loop, corresponding to Eq. (3):

$$\phi_n \leftarrow \phi - \beta^{meta} \cdot g_n(\phi^*), \quad (8)$$

$$\text{where } g_n(\phi^*) = \frac{\partial}{\partial \phi}\big[\mathcal{L}_{sim}^*(\phi) + \alpha^{def}\mathcal{L}_{def}^*(\phi)\big](\phi^*), \quad (9)$$

and $\beta^{meta}$ is the meta-learning rate. With gradient-based meta-learning methods, such as MAML [32], the cross-task meta-gradient $g_n(\phi^*)$ is computed directly to obtain the Jacobian for updating parameters, at the inner-loop-optimized weight values $\phi^*$. However, estimating the Jacobian involves computationally problematic second derivates; First-Order MAML [32] and Reptile [33] have been proposed to approximate this meta update step, and we adapt such approximations to train the interactive registration network.

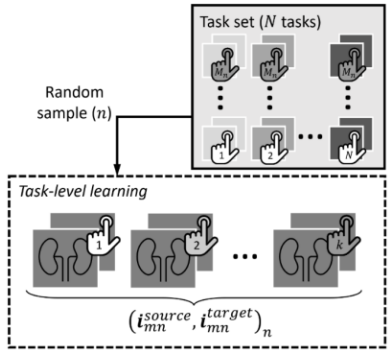

Fig. 1. Schematic of one episode of task-level learning. For each of the $k$ tasks in the sampled task set, the image pair is coupled with an associated annotation.

In the meta-test phase, parameters $\tilde{\phi}$ are adapted to the test task through few-shot learning. During meta-testing, a few interactions $\{i_{m,test}^{source}\}$ and $\{i_{m,test}^{target}\}$ are acquired from the test task to compute a few steps of test-task-specific gradients and update the model, using Eq. (6), before predicting the transformation using images $x_{test}^{source}$ and $x_{test}^{target}$ (Fig. 2).

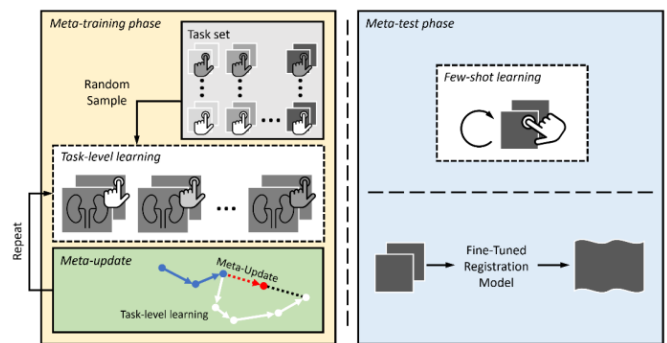

Fig. 2. An interactive meta-learning medical image registration framework. A learning-based registration model is trained over multiple episodes in meta-training (left) to learn an initialization for adaptation at inference. In each task-level learning episode, a task is sampled to train the model. Then, the meta-update (red arrow) updates the model based on the direction (black dashed line) of the task-level learning gradients (white arrows), continued from previously learned gradients (blue arrows). Later, the model is fine-tuned in the meta-test phase (right) with few-shot learning, coupled with user-defined interactions.

### D. Exemplar Clinical Application: Volume-to-Sparse Weakly-Supervised Multimodal Image Registration

In this section, we discuss our proposed methods for interactive registration to a real-world clinical application, in which only sparse TRUS images are available to be registered to preoperative MR images, using an interactive weakly-supervised multimodal image registration.

Prostate MR-TRUS image registration leverages MR imaging to aid tumour-targeted needle biopsy [40-47] and focal treatments [48-49] for suspected clinically significant prostate cancer. Image registration allows the presentation of MR-visible information, such as tumour size and location, for guiding surgical instruments or therapeutic energy placement. Often, the MR-derived lesion and tumour information are superimposed on the TRUS images as a visual aid.

A weakly-supervised methodology used to train an interactive registration network with a label-driven loss can be considered as a meta-learning problem, as described in Eq. (3) and Eq. (4), with $\alpha^{image} = 0$, without using explicit intensity-based similarity measures which have been considered less effective [1]. To accommodate sparse ultrasound images, readily available as interactions in this application, we develop a volume-to-sparse registration algorithm, where the training target images being a set of TRUS slices $\{x_{mn}^{target}\}$ and annotations of anatomical structures identified on these slices $\{l_{mn}^{target}\}$, with source MR images $\{x_{mn}^{source}\}$ and the corresponding MR annotations $\{l_{mn}^{source}\}$. These annotations can contain multiple types of anatomical structures [1], though this notation is omitted for brevity. We discuss the detailed representation of the interactive data in Section III.E and the need for TRUS slice localization information in the Discussion.

Our implementation utilizes LocalNet, a recent method for weakly-supervised image registration [1]. LocalNet's encoder-decoder structure comprises down- and up-sampling blocks and can predict a DDF that is summed over multiple resolutions. LocalNet is similar to the UNet [50] architecture found in VoxelMorph [2] – often used for unsupervised and weakly-supervised image registration. Compared to VoxelMorph, LocalNet has a smaller memory requirement and is more densely connected, with multiple types of residual shortcuts and summation-based skip layers to allow deeper supervision [1].

### E. Interactive Acquisition of Labelled TRUS Images

This study investigates an MR-TRUS registration where volume-to-sparse registration continually re-occurs throughout acquisition, as opposed to discreate registration to reconstructed 3D TRUS volumes. The continuous 2D TRUS images in such registration are considered the addition of new data, with or without the automatically acquired prostate gland segmentation [51], as interactions. At inference, this continuous stream of interactively acquired data provides additional context and a constantly up to date registration.

Here, interaction stems from the continual acquisition of frames by moving TRUS probe. Therefore, during few-shot learning, new frames are incorporated into the input of the model. This requires knowledge of the spatial relationship between each frame, so that the new frame may be inserted into the correct location within the TRUS volume. To provide initial spatial information for the network, the first interaction comprises two frames, and subsequent interactions require at least one new frame. Given the current clinical workflow for tumour-targeted needle biopsies, this interaction is unlikely to introduce any delay or modifications to existing protocols.

To simulate interactions in training, we select one pair of target interactions $i_{mn}^{target}$ by randomly selecting a series of TRUS images in a clinically feasible manner, whilst the target "interaction" is the fixed MR images and their annotation $i_{mn}^{source}$, as described in Scenario D in Section III.A. The label pair $\ell_{mn}^{source}$ and $\ell_{mn}^{target}$ may define either the prostate boundary, the apex and base of the prostate, or any other patient-specific landmarks; such as zonal structures, water-filled cysts, and calcifications [52]. The binary mask is generated to randomly include some number of frames $F$, where $F \in \mathbb{N} : F \in [F_{min}, F_{max}]$, which defines the image slices within the TRUS volume $x_{mn}^{target}$. Once generated, sections of the input image $x_{mn}^{target}$, and corresponding label $\ell_{mn}^{target}$ are masked-out, leaving only TRUS slices and corresponding labels from the simulated acquisition.

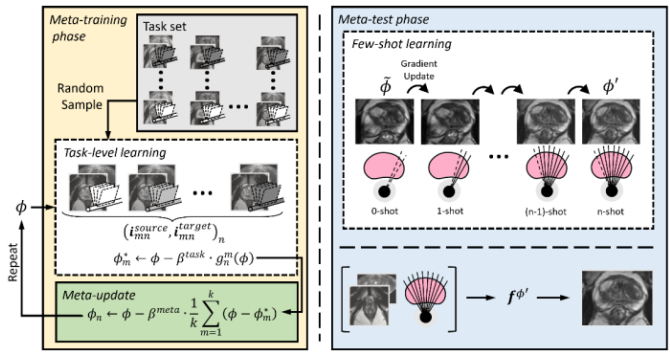

Fig. 3. Proposed framework for interactive medical image registration with meta-learning, as applied to weakly-supervised volume-to-sparse prostate MR-TRUS registration. The learner is trained over multiple episodes in meta-training (left) to learn an initialization for adaptation at inference. In each task-level learning episode, a set of images, labels, and some number of frames $F$ is sampled and trained on. After each episode, the meta-update updates the learner using the Reptile algorithm based on the task-level learning gradients. Once training is complete, the learner is optimized in the meta-test phase (right). Here, interactively-acquired data is coupled with few-shot learning to fine-tune a registration model in real-time as the TRUS image acquisition occurs.

### F. Meta-Learning an Initialization with Reptile

We adopt Reptile [33] as our gradient-based meta-learning strategy. Reptile provides a computationally efficient optimization of the gradient-based update procedure to approximate Eq. (8) and Eq. (9) by:

$$\phi_n \leftarrow \phi - \beta^{meta} \cdot \frac{1}{k} \sum_{m=1}^{k}(\phi - \phi_m^*), \quad (10)$$

where $\phi_m^*$ can be estimated using Eq. (6).

It is of note that, given that the complete prostate (and other patient-specific landmarks) labels are available, a stronger form of supervision is used to compute the loss during meta-training, such that the entire label similarity is computed rather than a partial similarity on sparse labels. This allows the initialization to be learned from complete data, illustrating how interactive labels and images used in computing training losses may differ from those seen in meta-testing in order to better guide learning.

During the meta-test phase, for evaluation, few-shot learning occurs with $F$ gradient updates on interactions $x_{mn}^{target}$ and $\ell_{mn}^{target}$ from the test task. This fine-tunes the model to obtain adapted parameters $\phi'$ which can perform accurate registrations on the test patient. Unlike the random generation of interactions during the meta-training phase, $x_{mn}^{target}$ and $\ell_{mn}^{target}$ define a continuous, single-sweep TRUS acquisition. Therefore, the first few-shot learning gradient update contains $F_{min}$ images and subsequent updates add an image, until the final update with $F_{max} - 1$ images. This ensures that the inference step is computed on an input with $F_{max}$ images. During the meta-test phase, we only use the label which defines the prostate boundary. This is done to emulate the labels which may be available (via automatic segmentation) in practice with the application of our method. A visual summary of the meta-learning phases for our application is shown in Fig. 3.

### G. Loss Functions

Two loss functions are used in training. In weakly-supervised registration, we seek to maximize the expected label similarity using a multiscale soft probabilistic Dice [1], which has shown effectiveness, especially when small foreground labels do not overlap initially. Using interactively acquired TRUS labels $\ell_{mn}^{target}$ and pre-operative MR labels $\ell_{mn}^{source}$, we obtain:

$$\mathcal{L}_{sim}^*(\phi) = \frac{1}{Z} \sum_\sigma \mathcal{S}_{Dice}\left(f_\sigma(\ell_{mn}^{target}), f_\sigma(\ell_{mn}^{source}(u_n^\phi))\right), \quad (11)$$

where $\mathcal{S}_{Dice}$ is the soft probabilistic Dice [53], $f_\sigma$ is a 3D Gaussian filter with an isotropic standard deviation $\sigma \in \{0, 1, 2, 4, 8, 16, 32\}$ in mm, and $Z = |\sigma|$. We additionally use bending energy [54] to regularize deformation $\mathcal{L}_{def}^*(\phi)$ on $u_n^\phi$ in tandem with $\mathcal{L}_{sim}^*(\phi)$ as in Eq. (3) and Eq. (4).

### H. Data

We used 108 pairs of pre-operative T2-weighted MR and intraoperative TRUS images from 76 patients, acquired during the SmartTarget clinical trials [52], a study approved by the London-Dulwich Research Ethics Committee (REF 14/LO/0830) and conducted at University College London Hospital (UCLH). Images were split into training and test sets comprising 88 and 20 images, respectively. No patient appears in both sets. Images were normalized and resampled to an isotropic voxel size of $0.8 \times 0.8 \times 0.8$ mm$^3$. MR segmentations were acquired as part of the SmartTarget clinical trial protocols [52]. TRUS prostate gland segmentations were acquired automatically [51], and landmarks were manually segmented.

### I. Baseline Model Implementation and Training

The framework was implemented in TensorFlow [55] and Keras [56]. The weakly-supervised registration framework and loss functions were adapted from DeepReg [57]. Hyper-parameters are as described in [1] unless otherwise specified. A random affine transformation, without flipping, was applied to each image-label pair for data augmentation.

The Baseline interactive registration model was trained for 250000 iterations with the Adam optimizer [58], a minibatch size of 4, and an initial learning rate, $\beta^{task}$, of $1 \times 10^{-5}$. In the meta-training phase, the value of $k$ for task-level learning was 10, and the initial meta-learning rate, $\beta^{meta}$, was set to 0.5, with a linear decay to $1 \times 10^{-5}$ at the final training iteration. Loss weights $\gamma$ and $\alpha$ were both set to 1.0. We let $F_{min} = 2$ and $F_{max} = 10$. Training took approximately 120 hours using one Tesla V100 GPU. We note that the number of iterations comprises each episode of task-level training, but does not include the meta-update; such is to say that we perform 25000 episodes of task-level learning, where each episode of task-level learning encompasses $k$ gradient updates.

### J. Comparison with Meta-Learning Variants

Without extensively searching all hyper-parameters, which may misrepresent generalizability, we provide experimental results and validation on variants to the Baseline. First, we modify the number of gradient updates performed in task-level learning, $k$, to 1 and 100. Notably, when $k = 1$, a single step of SGD on the expected loss is equivalent to jointly training on a mixture of all tasks [33]. Though $k$ is often defined as $\leq 10$ in other meta-learning applications [33], we also demonstrate training with a higher value. Due to the changes introduced to the training process (for $k = 1$), and the deviation of the gradients from those which would normally be encountered in a non-meta-learning-based training protocol (for $k = 100$), these variants will likely underperform relative to the baseline. Second, we modify the initial meta-learning rate, $\beta^{meta}$, to 0.25 and 1.0. The linear decay remains unchanged. To prevent arbitrary selection, we choose values corresponding closely to those presented in [33]. Finally, we vary the maximum number of frames used in training, $F_{max}$, to 5 and 15. We expect a higher and lower $F_{max}$ would result in better and worse performance, respectively. Though if the increase in performance gained per frame diminishes as $F_{max}$ increases, training with a smaller $F_{max}$ may be beneficial. Conversely, if the increase in performance per frame does not significantly diminish, training with a higher $F_{max}$ and acquiring additional frames in practice may be prudent.

### K. Comparison with State-of-the-Art Approaches

We compare the proposed Baseline to the application of 'registration' without alignment, and of a simple initialization whereby the prostate gland centroids are aligned. Furthermore, we compare to two weakly-supervised state-of-the-art approaches for deformable pairwise medical image registration; LocalNet [1], and VoxelMorph [2]. In all comparisons, we use complete 3D volumes for source and target input images – unlike our interactive meta-learning approach which provides a sparse target input. Hyper-parameters are all kept at defaults as described in [1] and [2], and we set loss weights $\gamma$ and $\alpha$ to 1.0.

### L. Comparison with Non-Meta Learning Approaches

We emulate the sparse 2D target input of our interactive meta-learning approach with instances of LocalNet and VoxelMorph with 5 or 10 randomly sampled 2D target input images in training. We demonstrate the effects of few-shot learning on these models trained without meta-optimization and our meta-learning Baseline by performing inference with and without any few-shot learning. To illustrate the effectiveness of the meta-learned initialization, we randomly initialize LocalNet and VoxelMorph models and apply few-shot learning to the networks. While the impact of sparse data was not investigated in [1] or [2], and therefore, may adversely impact their performance, this assessment provides a benchmark to which we may compare the performance of our approach to learning-based methods with comparable amounts of input data.

### M. Evaluation of Registration Methods

To compare the Baseline to all aforementioned methods, we test interactions which represent a clinically realistic scenario on our real-world, clinical test data. Through the continuous acquisition of frames from a right-to-left sweep through the prostate, we obtain a series of sagittal images which are uniformly distributed through the prostate (Fig. 4). As noted in Section III.F, we initially acquire two images (as $F_{min} = 2$) to provide spatial context of the frames in this first acquisition.

Registration accuracy was quantified using the Dice similarity coefficient (DSC) and target registration error (TRE). Two-tailed paired t-tests, at significance level α = 0.05, are used to compare each method to the Baseline. DSC is computed between the warped MR label and the entire ground-truth TRUS label. TRE is defined as the root-mean-square distance between the geometric centroids of the registered landmark pairs. In our dataset, landmarks consisted of 309 pairs of manually identified points, including the apex and base of the prostate, and various patient-specific landmarks including zonal boundaries, water-filled cysts, and calcifications [1, 52, 59]. Notably, such landmarks have been previously utilized yield an overall spatial distribution which is representative of the full TRE distribution in this application [1, 4-5, 7-8, 10-11, 13, 60-70], therefore providing an evaluation of registration accuracy and an estimate of registration errors, such as those associated with tumour localization. We also report the computational time per few-shot learning gradient update and subsequent registration in the meta-test phase for our approach.

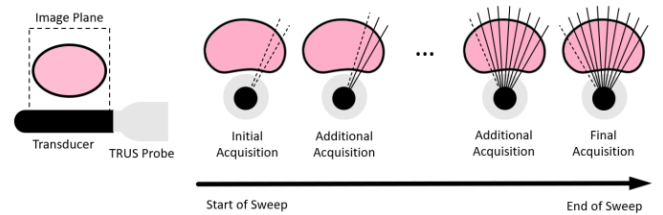

Fig. 4. Illustration of TRUS images acquired in the presented clinical scenario. Acquired images (dashed lines) are captured in the sagittal plane (left) and shown with previously acquired images (solid lines) through one continuous 'sweep' of the prostate with the TRUS probe until full coverage is obtained.

## IV. RESULTS

### A. Baseline Performance

One gradient update and inference for the Baseline model requires 0.67 ± 0.07s and 0.37 ± 0.05s, respectively. Therefore, during adaptation, which may occur during image acquisition, approximately 6s is needed to perform fine-tuning and inference, considerably faster than the 2-4 mins required for acquisition, contouring, and registration in conventional image-fusion targeted biopsies, such as those reported in the SmartTarget clinical trials [52].

After few-shot learning, we achieved a median TRE of 4.26 mm and a mean DSC of 0.85 with 10 input TRUS frames. This is within range of previously defined clinically significant thresholds of 2.97 mm [71] and 5.00 mm [62]. A detailed summary of TRE and DSC is given in Table 1. Example slices of input MR and TRUS image pairs and registered MR images are provided in Fig. 5 for qualitative visual assessment for the Baseline through each few-shot step in the meta-test phase.

TABLE I
SUMMARY TRE AND DSC FOR THE BASELINE NETWORK AT EACH STEP OF FEW-SHOT LEARNING. VALUES ARE PRESENTED ± SD. TRE IN MM.

| $F$ | Grad. Updates | Median TRE | Mean DSC |
|---|---|---|---|
| 2 | 0 | 7.02 ± 4.08 | 0.77 ± 0.06 |
| 3 | 1 | 6.98 ± 3.98 | 0.79 ± 0.06 |
| 4 | 2 | 6.02 ± 4.17 | 0.81 ± 0.06 |
| 5 | 3 | 5.61 ± 4.11 | 0.82 ± 0.07 |
| 6 | 4 | 5.34 ± 4.16 | 0.82 ± 0.07 |
| 7 | 5 | 5.27 ± 4.08 | 0.83 ± 0.07 |
| 8 | 6 | 4.34 ± 4.12 | 0.84 ± 0.06 |
| 9 | 7 | 4.37 ± 4.13 | 0.84 ± 0.06 |
| 10 | 8 | 4.26 ± 4.19 | 0.85 ± 0.06 |

### B. Performance of Baseline Variants

After few-shot learning, the $k = 1$ variant had a median TRE of 4.48 mm and mean DSC of 0.83, whereas the $k = 100$ variant had a median TRE of 4.58 mm and mean DSC of 0.85. In both, no significant difference was found between TRE or DSC relative to the Baseline. The effects of $k$ in training on TRE are illustrated in Fig. 6 and summarized in Table 2.

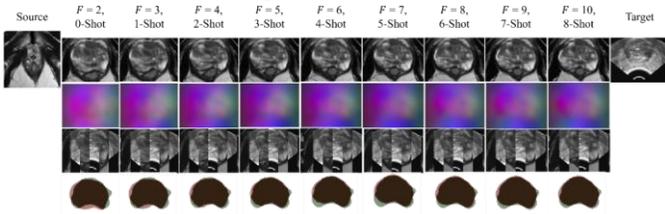

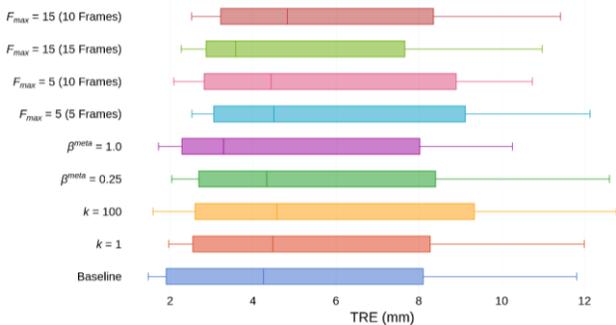

Fig. 5. Example image slice from one test case. The left-most column contains an image slice from source MR volume. The right-most column contains the corresponding target TRUS image slice. Other columns present the warped source MR image, resulting DDF, alternating vertical slices of the warped MR and target TRUS image, and warped MR prostate gland contour (Red) overlaid on the target TRUS prostate gland contour (Green), using the Baseline at a given shot of training during few-shot learning, with $F$ frames.

Fig. 6. Tukey's boxplots of TRE for the Baseline and all variants in the MR-TRUS registration experiment. Whiskers indicate 10th and 90th percentiles. Results are presented for registrations with 10 frames unless otherwise indicated. For $F_{max}$ variants, we additionally present results at 10 frames for direct comparison to the Baseline and other variants.

TABLE II
SUMMARY TRE AND DSC FOR THE $k = 1, 100$ VARIANTS AT EACH STEP OF FEW-SHOT LEARNING. VALUES ARE PRESENTED ± SD. TRE IN MM.

| $k$ | $F$ | Grad. Updates | Median TRE | Mean DSC |
|---|---|---|---|---|
| 1 | 2 | 0 | 7.79 ± 3.75 | 0.76 ± 0.06 |
| | 3 | 1 | 7.55 ± 3.88 | 0.78 ± 0.06 |
| | 4 | 2 | 7.03 ± 3.97 | 0.79 ± 0.06 |
| | 5 | 3 | 5.79 ± 3.90 | 0.80 ± 0.06 |
| | 6 | 4 | 5.59 ± 3.90 | 0.80 ± 0.06 |
| | 7 | 5 | 5.49 ± 4.12 | 0.80 ± 0.06 |
| | 8 | 6 | 4.93 ± 4.07 | 0.81 ± 0.06 |
| | 9 | 7 | 4.43 ± 4.01 | 0.83 ± 0.06 |
| | 10 | 8 | 4.48 ± 3.96 | 0.83 ± 0.05 |
| 100 | 2 | 0 | 7.83 ± 3.86 | 0.76 ± 0.06 |
| | 3 | 1 | 7.05 ± 4.00 | 0.78 ± 0.06 |
| | 4 | 2 | 6.49 ± 4.12 | 0.79 ± 0.06 |
| | 5 | 3 | 5.88 ± 4.20 | 0.81 ± 0.06 |
| | 6 | 4 | 6.03 ± 4.32 | 0.82 ± 0.06 |
| | 7 | 5 | 5.64 ± 4.43 | 0.83 ± 0.06 |
| | 8 | 6 | 5.18 ± 4.44 | 0.84 ± 0.05 |
| | 9 | 7 | 4.78 ± 4.48 | 0.84 ± 0.05 |
| | 10 | 8 | 4.58 ± 4.48 | 0.85 ± 0.04 |

After few-shot learning, the $\beta^{meta} = 0.25$ variant had a median TRE of 4.33 mm and a mean DSC of 0.84, whereas the $\beta^{meta} = 1.0$ had a median TRE of 3.29 mm and a mean DSC of 0.87, no significant difference to the Baseline, as summarized in Table 3 and Fig. 6 with varying $\beta^{meta}$.

After $F_{max} - F_{min}$ gradient updates of few-shot learning, the $F_{max} = 5$ variant had a median TRE of 4.50 mm and mean DSC of 0.85, whereas the $F_{max} = 15$ variant had a median TRE of 3.58 mm and mean DSC of 0.84, no significant difference to the Baseline. The effects of $F_{max}$ during training on TRE are illustrated in Fig. 6, and given for TRE and DSC in Table 4.

We note that the $F_{max} = 5$ variant performs better than the $F_{max} = 15$ variant for all values of $F \leq 5$. This is likely due to the distribution of the input images in the presented clinical scenario, whereby one continuous sweep of the prostate occurs, as presented in Fig 4. For example, when $F = 5$, while the input frames of the $F_{max} = 5$ variant will be evenly distributed across the entire prostate, while the 5 input frames of the $F_{max} = 15$ variant will be condensed into the right-most third of the prostate, resulting in less spatial information being presented about the remaining prostate volume.

Example slices of input MR and TRUS image pairs and registered MR images are provided in Fig. 7 for qualitative visual assessment of the results for each variant.

TABLE III
SUMMARY TRE AND DSC FOR THE $\beta^{meta} = 0.25, 1.0$ VARIANTS AT EACH STEP OF FEW-SHOT LEARNING. VALUES ARE PRESENTED ± SD. TRE IN MM.

| $\beta^{meta}$ | $F$ | Grad. Updates | Median TRE | Mean DSC |
|---|---|---|---|---|
| 0.25 | 2 | 0 | 7.06 ± 4.00 | 0.75 ± 0.07 |
|  | 3 | 1 | 6.95 ± 3.95 | 0.76 ± 0.07 |
|  | 4 | 2 | 6.44 ± 4.03 | 0.78 ± 0.06 |
|  | 5 | 3 | 5.70 ± 3.88 | 0.80 ± 0.05 |
|  | 6 | 4 | 5.35 ± 4.00 | 0.80 ± 0.05 |
|  | 7 | 5 | 5.26 ± 4.04 | 0.81 ± 0.05 |
|  | 8 | 6 | 4.62 ± 4.06 | 0.82 ± 0.05 |
|  | 9 | 7 | 4.31 ± 4.11 | 0.83 ± 0.05 |
|  | 10 | 8 | 4.33 ± 4.11 | 0.84 ± 0.05 |
| 1.0 | 2 | 0 | 7.54 ± 3.76 | 0.79 ± 0.05 |
|  | 3 | 1 | 7.17 ± 3.77 | 0.81 ± 0.05 |
|  | 4 | 2 | 6.62 ± 3.73 | 0.83 ± 0.05 |
|  | 5 | 3 | 5.05 ± 3.64 | 0.84 ± 0.05 |
|  | 6 | 4 | 4.41 ± 3.64 | 0.84 ± 0.04 |
|  | 7 | 5 | 4.22 ± 3.71 | 0.85 ± 0.04 |
|  | 8 | 6 | 3.64 ± 3.84 | 0.86 ± 0.04 |
|  | 9 | 7 | 3.22 ± 3.93 | 0.87 ± 0.04 |
|  | 10 | 8 | 3.29 ± 3.97 | 0.87 ± 0.04 |

TABLE IV
SUMMARY TRE AND DSC FOR THE $F_{max} = 5, 15$ VARIANTS AT EACH STEP OF FEW-SHOT LEARNING. VALUES ARE PRESENTED ± SD. TRE IN MM.

| $F_{max}$ | $F$ | Grad. Updates | Median TRE | Mean DSC |
|---|---|---|---|---|
| 5 | 2 | 0 | 6.49 ± 3.80 | 0.79 ± 0.06 |
|  | 3 | 1 | 5.67 ± 3.94 | 0.82 ± 0.06 |
|  | 4 | 2 | 4.58 ± 3.91 | 0.84 ± 0.05 |
|  | 5 | 3 | 4.50 ± 3.93 | 0.85 ± 0.04 |
| 15 | 2 | 0 | 7.19 ± 4.01 | 0.76 ± 0.04 |
|  | 3 | 1 | 6.82 ± 4.08 | 0.77 ± 0.05 |
|  | 4 | 2 | 6.53 ± 4.19 | 0.78 ± 0.05 |
|  | 5 | 3 | 6.33 ± 4.27 | 0.79 ± 0.05 |
|  | 6 | 4 | 5.71 ± 4.16 | 0.80 ± 0.05 |
|  | 7 | 5 | 5.51 ± 4.14 | 0.81 ± 0.06 |
|  | 8 | 6 | 5.36 ± 4.10 | 0.81 ± 0.06 |
|  | 9 | 7 | 5.44 ± 4.06 | 0.81 ± 0.06 |
|  | 10 | 8 | 4.83 ± 4.03 | 0.81 ± 0.06 |
|  | 11 | 9 | 4.37 ± 3.99 | 0.82 ± 0.06 |
|  | 12 | 10 | 4.03 ± 3.96 | 0.83 ± 0.06 |
|  | 13 | 11 | 3.86 ± 3.95 | 0.84 ± 0.06 |
|  | 14 | 12 | 3.64 ± 4.01 | 0.84 ± 0.06 |
|  | 15 | 13 | 3.58 ± 4.00 | 0.84 ± 0.06 |

### C. Performance of State-of-the-Art Approaches

With no initial registration or alignment, a median TRE of 32.4 mm and mean DSC of 0.66 are obtained. Further, a median TRE of 18.4 mm and mean DSC of 0.77 are obtained if only prostate gland centroid alignment is performed.

The performance of the Baseline model was not significantly different than LocalNet for TRE and DSC, where a median TRE and mean DSC of 3.97 mm and 0.87 are obtained. The performance of the Baseline model was not significantly different than VoxelMorph, for TRE and DSC, where a median TRE and mean DSC of 4.32 mm and 0.84 are obtained.

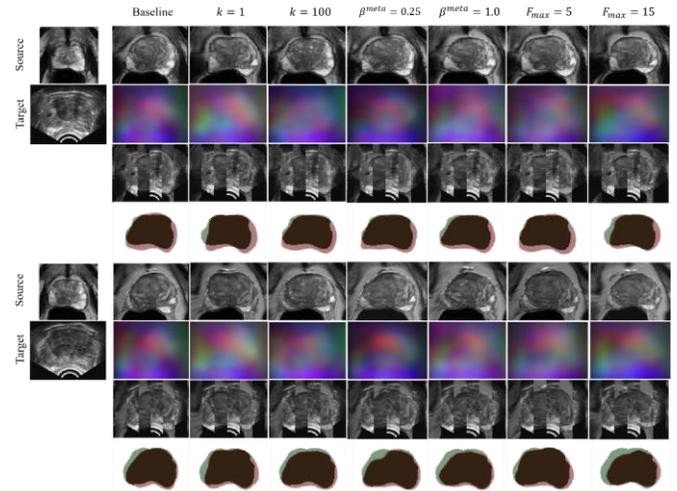

Fig. 7. Example image slices from one test case. The left-most column contains image slices from source MR volume and corresponding target TRUS image slice. Other columns present the warped source MR image, resulting DDF, alternating vertical slices of the warped MR and target TRUS image, and warped MR prostate gland contour (Red) overlaid on the target TRUS prostate gland contour (Green), using the above-labelled network.

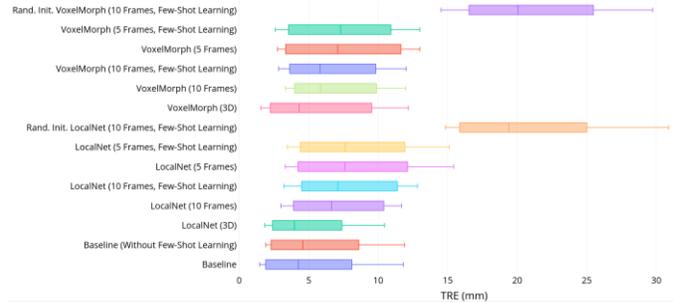

Fig. 8. Tukey's boxplots of TRE for Baseline, state-of-the-art, and all non-meta-learning methods in the MR-TRUS registration experiment. Whiskers indicate 10th and 90th percentiles. Baseline results presented for registrations with 10 frames, with input size indicated explicitly for all other methods.

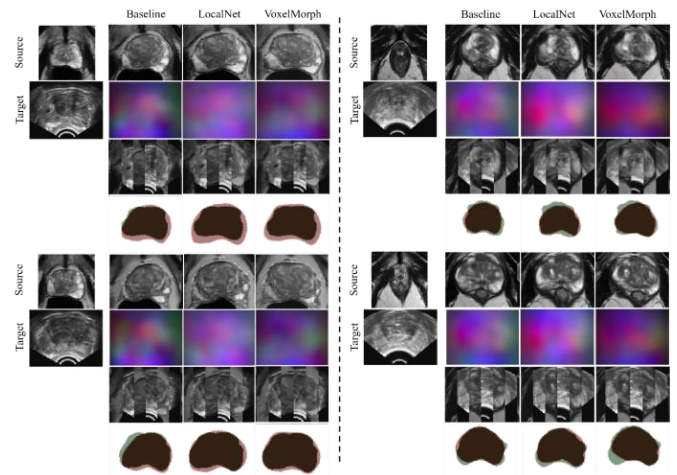

Fig. 9. Example image slices from two test cases. The left-most column contains image slices from source MR volume and the corresponding target TRUS image slice. Other columns present the warped source MR image, resulting DDF, alternating vertical slices of the warped MR and target TRUS image, and warped MR prostate gland contour (Red) overlaid on the target TRUS prostate gland contour (Green), using the above-labelled network.

A summary of TRE for the Baseline and the non-meta-learning-based methods is given in Fig. 8. Example slices of input MR and TRUS image pairs and the registered MR images are provided in Fig. 9 for qualitative visual assessment of the registration results for each approach. It is important to note that these methods use complete 3D volumes for source and target input images, and only achieve comparable performance to our method, which uses between two and ten frames of the target image in training and at inference. This represents between 1.6% and 8.5% of the complete 3D volume, which contains 118 image slices.

### D. Performance of Non-Meta-Learning Approaches

When emulating sparse input on LocalNet, a median TRE of 7.51 mm and a mean DSC of 0.76 are obtained with 5 input images. A median TRE of 6.26 mm and mean DSC of 0.79 are obtained with 10 input images. Fine-tuned Baseline model performance is significantly different than when providing 5 ($p < 0.01$) and 10 ($p = 0.04$) input images for TRE. No significant difference is observed for DSC. When using VoxelMorph, a median TRE of 7.36 mm and a mean DSC of 0.78 are obtained with 5 input images. A median TRE of 5.86 mm and mean DSC of 0.81 are obtained with 10 input images. Performance of the fine-tuned Baseline model is significantly different than when providing 5 input images ($p < 0.01$), but not 10 images, for TRE. No significant difference is observed for DSC.

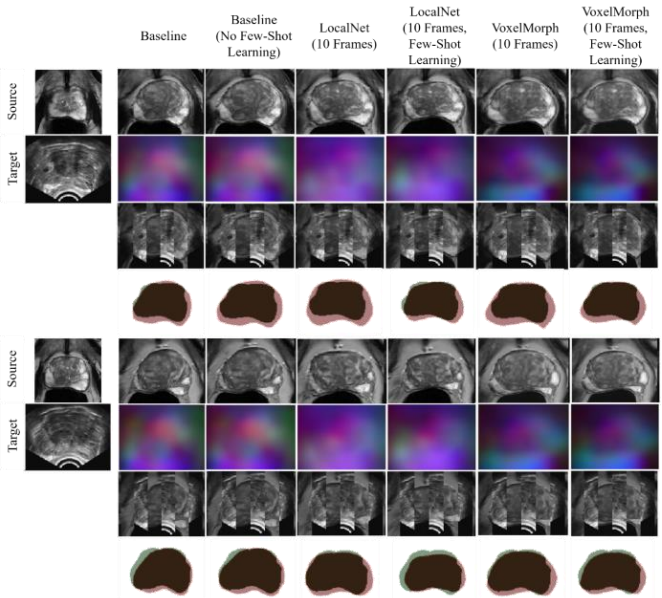

Fig. 10. Example image slices from one test case. The left-most column contains image slices from source MR volume and corresponding target TRUS image slice. Other columns present the warped source MR image, resulting DDF, alternating vertical slices of the warped MR and target TRUS image, and warped MR prostate gland contour (Red) overlaid on the target TRUS prostate gland contour (Green), using the above-labelled network.

Applying few-shot learning to these same models at inference, LocalNet obtains a median TRE of 7.64 mm and mean DSC of 0.76 are obtained with 5 input images. A median TRE of 7.23 mm and mean DSC of 0.73 are obtained with 10 input images. VoxelMorph obtains a median TRE of 7.30 mm and mean DSC of 0.79 are obtained with 5 input images. A median TRE of 5.81 mm and mean DSC of 0.81 are obtained with 10 input images. This suggests that few-shot learning has little effect when applied to conventionally trained registration networks, without the meta-trained network initialization.

Using the Baseline, without few-shot learning, a higher median TRE of 4.57 mm and a lower mean DSC of 0.82 is obtained without detected significance, compared to the Baseline when using few-shot learning. Applying few-shot learning to an untrained model, where the weights are initialized randomly, results in a median TRE of 19.4 mm and a mean DSC of 0.76 for LocalNet, and a median TRE of 20.1 mm and a mean DSC of 0.77 for VoxelMorph.

Detailed results summarizing the TRE of the Baseline and the non-meta-learning-based methods are illustrated in Fig. 8. Example slices of input MR and TRUS image pairs and the registered MR images are provided in Fig. 10 for qualitative visual assessment of the registration results for each approach.

## V. DISCUSSION

This work presents a deep learning framework for interactive medical image registration using meta-learning. As illustrated in Fig. 9, the performance of our Baseline network for volume-to-sparse registration provides accuracy that is comparable from recent 3D-to-3D methods, while using a fraction of the data. Further, it yields significantly improved metrics compared to other tested volume-to-sparse methods and indicates that our method is not sensitive to meta-learning hyper-parameters, demonstrating flexibility and generalizability. This motivates use for other registration applications.

Of importance for multimodal image registration, the lack of robust voxel-level similarity between image pairs necessitates the tested weakly-supervised registration algorithms, which require labelled structures in training, but not at inference. As we utilize few-shot learning in the meta-test phase, real-time prostate segmentations may be required on 2D TRUS images. High DSC and rapid inference times have been reported for this task [13, 51], as such, the need for segmentation must be considered, but should not be considered prohibitive to the real-time implementation of interactive registration in practice given that the addition of these additional segmentation inference steps would add, at most, several seconds to the total time required to compute the registration.

All employed volume-to-sparse methods require positional information for the TRUS images relative to a fixed reference. In practice, this may be obtained using positional, mechanical, or electromagnetic/optical tracking. Assessing our method's suitability for un-tracked TRUS images, however, is considered out of the scope of this work.

## VI. CONCLUSION

This paper presents a novel interactive image registration approach, using an exemplar application of partial registration of MR to sparsely acquired intra-operative TRUS images. We obtain similar registration accuracies to state-of-the-art 3D image registration methods which require complete image volumes. Our method significantly outperforms alternative methods when applied to the same challenging partial data problem. This work demonstrates the effectiveness and efficiency of our real-time interactive image registration method, which may be applied during intraoperative procedures, such as prostate biopsy.